\documentclass{INTERSPEECH2023}
\usepackage{tabularx}
\usepackage{booktabs}
\usepackage{tabularray}
\usepackage{xcolor}
\usepackage{multirow}
\usepackage{hyperref}
\usepackage[flushleft]{threeparttable}
\makeatletter
\renewcommand{\section}{\@startsection
  {section}%
  {1}%
  {}%
  {0.25\baselineskip}%
  {0.25\baselineskip}%
  {}}%

\renewcommand{\subsection}{\@startsection
  {subsection}%
  {2}%
  {}%
  {0.2\baselineskip}%
  {0.2\baselineskip}%
  {}}%

\renewcommand{\subsubsection}{\@startsection
  {subsubsection}%
  {3}%
  {}%
  {0.1\baselineskip}%
  {0.1\baselineskip}%
  {}}%
\let\oldtexttt\texttt 
\renewcommand{\texttt}[1]{{\footnotesize\oldtexttt{#1}}} 

\makeatother

\title{The CHiME-7 Challenge: System Description and Performance of \\ NeMo Team's DASR System}
\name{Tae Jin Park, He Huang, Ante Juki\'{c}, Kunal Dhawan, Krishna C. Puvvada, Nithin Koluguri, Nikolay Karpov, Aleksandr Laptev, Jagadeesh Balam and Boris Ginsburg}
\address{NVIDIA, Santa Clara, USA}
\email{\{taejinp,heh,ajukic,kdhawan,kpuvvada,nkoluguri,nkarpov,alaptev,jbalam,bginsburg\}@nvidia.com}

\begin{document}

\maketitle
 
\begin{abstract}
We present the NVIDIA NeMo team's multi-channel speech recognition system for the 7th CHiME Challenge Distant Automatic Speech Recognition (DASR) Task, focusing on the development of a multi-channel, multi-speaker speech recognition system tailored to transcribe speech from distributed microphones and microphone arrays. The system predominantly comprises of the following integral modules: the Speaker Diarization Module, Multi-channel Audio Front-End Processing Module, and the ASR Module. These components collectively establish a cascading system, meticulously processing multi-channel and multi-speaker audio input. Moreover, this paper highlights the comprehensive optimization process that significantly enhanced our system's performance. Our team's submission is largely based on NeMo toolkits and will be publicly available. 
\end{abstract}

\section{Introduction}
The landscape of conversational artificial intelligence (AI) has seen significant advancements in recent years, with an increased emphasis on applications such as automatic speech recognition (ASR), text-to-speech synthesis (TTS), and extensive use of large language models (LLMs). Among the numerous open-source toolkits in this domain, NVIDIA's NeMo is emerging as a pivotal toolkit for conversational AI, particularly ASR models. A hallmark of the NeMo toolkit is its ability to leverage prior work, in terms of both code and pretrained models, thereby enhancing the development and evaluation process more efficient and reproducible. Our submission for \textit{CHiME-7} DASR task is largely based on the NeMo models and code bases which are originally built for single-channel tasks. Most importantly, we publicly release our submission based on NeMo toolkit so that anyone could test and improve the proposed system.

In the context of practical applications, this paper explores NVIDIA NeMo team in the 7th CHiME Challenge.~Specifically, we shed light on the intricacies of their multi-channel speech recognition system, designed to transcribespeech picked up from multiple, distributed microphones and arrays. While our focus is honed in on Track 1, encompassing distant automatic speech recognition (DASR), the details of this system are revealed through its principal components, such as the speaker diarization module, the multi-channel (MC) Audio front-end processing module, and the ASR Module. Collectively, these modules create a cascade that adeptly handles multi-channel, multi-speaker audio input. We will also highlight the optimization methods employed to boost the system's efficiency, as showcased on the development set.

This paper is constructed in the following structure. In Section \ref{sec:proposed_dasr_system} we explain the details of each signal processing chain and neural models that comprise our DASR system. In Section \ref{sec:hyper_parameter_optimization} we discuss the hyper-parameter optimization method used to adjust the non-differentiable parameters affecting the overall performance. In Section \ref{sec:experimental_results}, we present the evaluation results on the dev and eval set of the challenge dataset. Finally, we conclude and discuss the future work in Section \ref{sec:conclusions}.

\begin{figure}[t]
\centering
\includegraphics[width=0.44\textwidth]{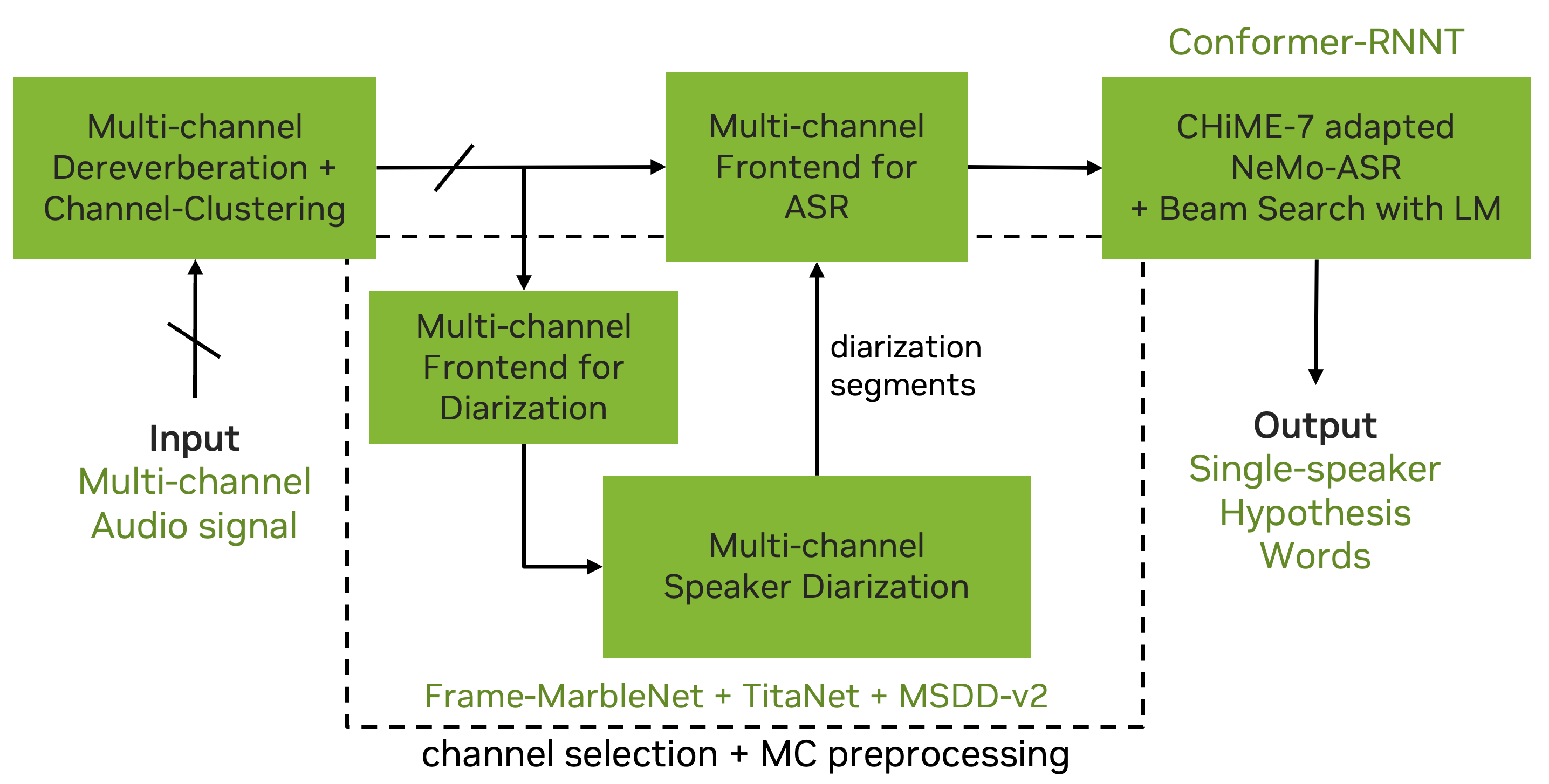}
\vspace{-6px}
\caption{Dataflow of the NeMo Team's CHiME-7 Submission.}
\vspace{-1px}
\label{fig:llm_ast_diar}
\end{figure}

\begin{figure}[t]
\centering
\includegraphics[width=0.44\textwidth]{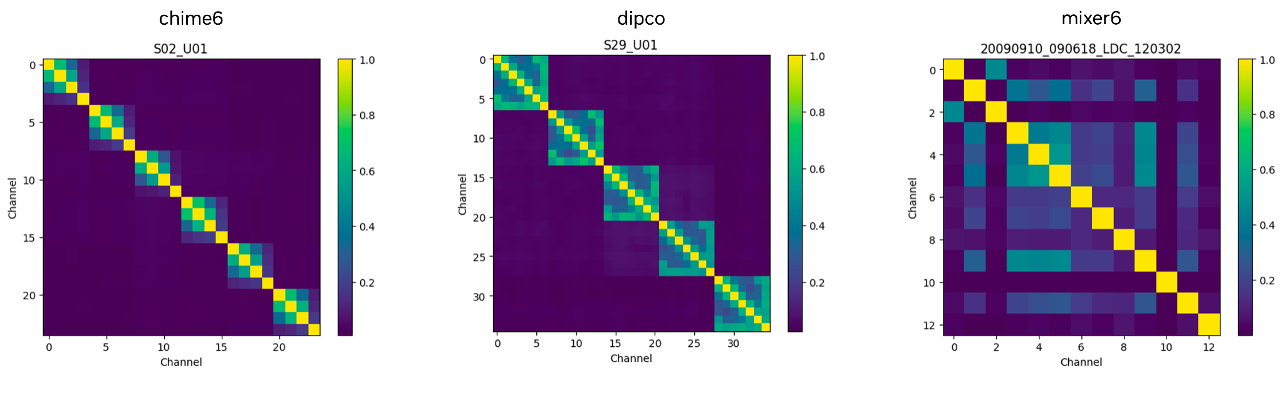}
\vspace{-6px}
\caption{Channel clustering: $\bar{\mathbf{\Gamma}}$ computed on a single session from CHiME-6 (left), DiPCo (middle), and Mixer 6 (right) development subsets.}
\vspace{-1px}
\label{fig:ch_clus}
\end{figure}


\section{Proposed DASR system}
\label{sec:proposed_dasr_system}

The Fig.~\ref{fig:mc_diar_dataflow} illustrates the data flow of our submission for the DASR task. In the highest level, our system consists of the following modules: dereverberation with channel clustering, front-end for diarization, speaker diarization module, front-end for ASR, ASR decoder module. 

\subsection{Dereverberation with channel-clustering}
\label{ssec:derev_with_ch_clus}
This module aims to perform dereverberation and channel clustering on the raw multi-channel audio input. The goal of the dereverberation module is to reduce the impact of room reverberation on diarization performance, similar to previous submissions~\cite{medennikov2020stc}. The objective of the channel clustering module is to decrease the number of audio streams utilized for multi-channel diarization.
The multi-channel audio signals are processed using the weighted prediction error-based dereverberation for multiple-input multiple-output (MIMO) as presented in~\cite{yoshioka_2012} and implemented in NeMo toolkit~\cite{nemo_2019}.
Block-wise MIMO dereverberation is conducted over 40-second windows with a two-second window overlap. This process utilizes STFT with 64~ms window length, 75\% overlap, 10 frame filters, a 3-frame prediction delay and 10 iterations.
Using the processed multi-channel signal, we compute a magnitude-squared coherence (MSC) matrix, $\mathbf{\Gamma}(f) \in \mathbb{R}^{M \times M}$, where $M$ represents the number of channels. This approach is consistent with the methodology in~\cite{coherence_clustering_2021}, and the elements in $\mathbf{\Gamma}(f)$ are:  
\begin{align}
\left\lbrace \mathbf{\Gamma}(f) \right\rbrace_{ij} = \frac{|S_{ij}(f)|^2}{S_{ii}(f) S_{jj}(f)},
\end{align}
where $S_{ij}(f)$ is the cross-power spectral density between channels $i$ and $j$.
The average MSC matrix $\bar{\mathbf{\Gamma}}$ is obtained by averaging over frequency subbands $f$ between 300~Hz and 3.5~kHz.
Fig.~\ref{fig:ch_clus} depicts examples of $\bar{\mathbf{\Gamma}}$ for a single session from each of the subsets of the development set.
It can be observed that the patterns in $\bar{\mathbf{\Gamma}}$ correspond to different microphone array configurations used for recording the data for each of the subsets.
Clustering of the channels is obtained by applying maximum eigengap spectral clustering (NME-SC)~\cite{park_2020} on $\bar{\mathbf{\Gamma}}$.
Note that the number of clusters is estimated automatically using normalized maximum eigengap~\cite{park_2020}, and channel clustering is performed for each session independently.
The obtained channel clusters are used to reduce the number of audio streams for multi-channel diarization.
Signals within each cluster are averaged, and these output streams are used as the input for the following steps of multi-channel speaker diarization.

\subsection{Speaker diarization}
\label{ssec:speaker_diarization}

\label{sec:mcp}

\subsubsection{Multi-channel Voice Activity Detection}

\begin{figure}[t]
\centering
\includegraphics[width=0.45\textwidth]{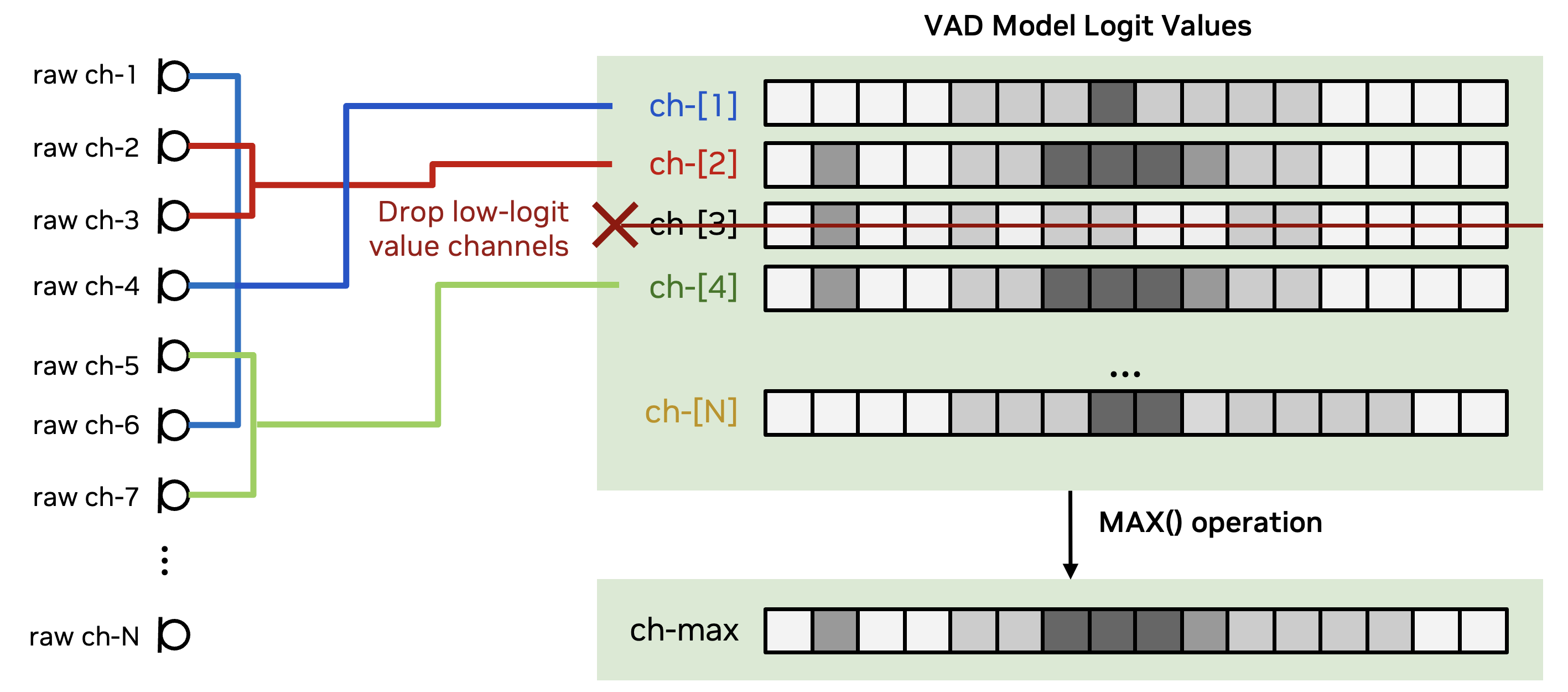}
\vspace{-3px}
\caption{Multi-channel VAD is applied on clustered channels while low-logit channels are dropped. Finally, frame-wise max pooling is applied across the remaining clustered channels.}
\vspace{3px}
\label{fig:mc_vad}
\end{figure}

\begin{figure}[t]
\centering
\includegraphics[width=0.45\textwidth]{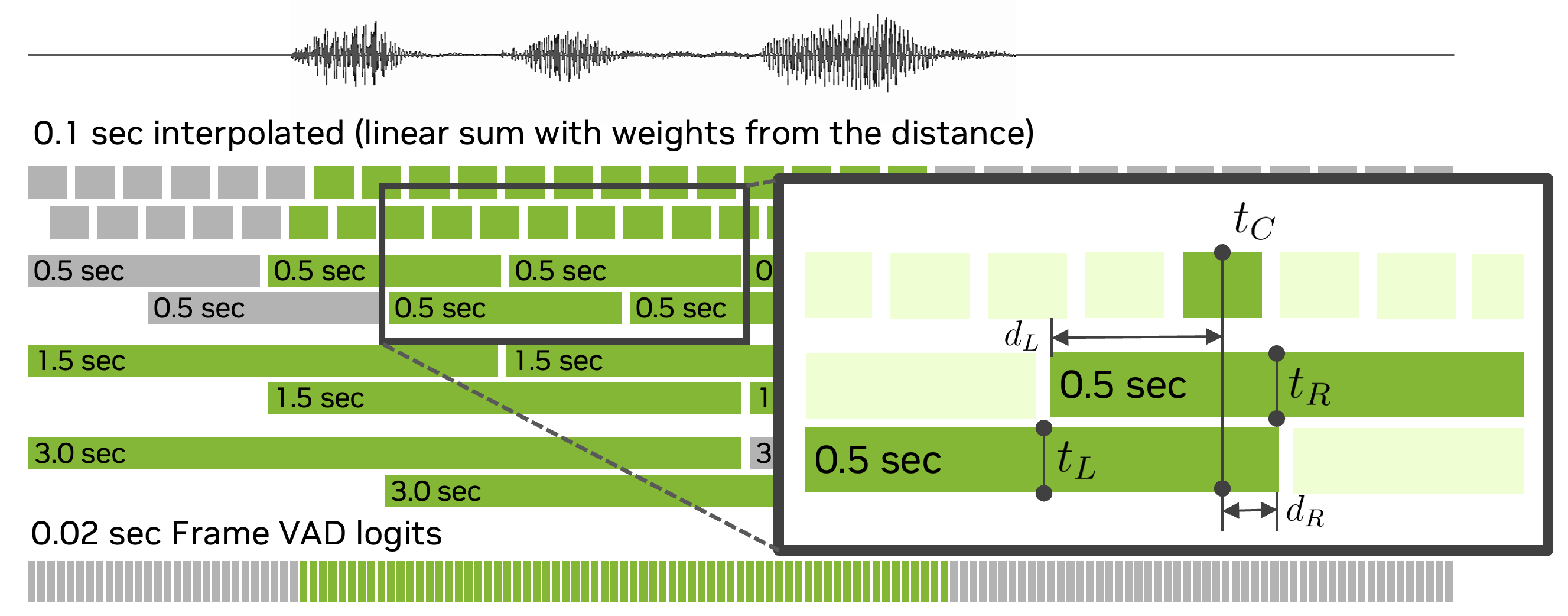}
\vspace{-3px}
\caption{Speech signal and its corresponding multi-scale segments for speaker embeddings. The frame-VAD logits will mask the only frames detected as speech.}
\vspace{3px}
\label{fig:frame_vad_mask}
\end{figure}

\label{sec:vad}
We employ a convolution-based voice activity detection (VAD) model to predict speech probability for each 20\,ms block of the input audio signal~\cite{speech_to_frame_label}.
The model is randomly initialized and trained on a combination of the \mbox{CHiME-6}~\cite{watanabe20b_chime} training subset and an additional simulated dataset.
For validation, we use dataset comprises the \mbox{CHiME-6} development subset as well as 50\,h of simulated audio data.
The simulated data is generated using the NeMo multi-speaker data simulator~\cite{speech_data_simulator} on VoxCeleb1\&2 datasets~\cite{nagrani2017voxceleb,chung2018voxceleb2}.This results in a total of 2,000 hours of audio, of which approximately 30\% is silence. Additionally, the model training incorporates SpecAugment~\cite{park2019specaugment} and noise augmentation through MUSAN~\cite{snyder2015musan}. During inference, the input to the VAD model first undergoes processing by the front-end system, as detailed in Section~\ref{sec:mcp}. For multi-channel VAD inference, we employ the VAD model trained on a single-channel signal.
We exclude channels that have relatively low average VAD probabilities, ranging from 0\% to 50\%. Subsequently, a maximum operation is applied to the VAD probabilities for each frame, generating the final VAD probability values for the multi-channel audio input.

\subsubsection{Multi-channel Diarization Module}
Our DASR system is based on the speaker diarization system using the multi-scale diarization decoder (MSDD) proposed in~\cite{park2022multi}. This system employs a multi-scale embedding approach and utilizes TitaNet~\cite{koluguri2022titanet} speaker embedding extractor. 
Unlike the system in~\cite{park2022multi} that uses a multi-layer long short-term memory (LSTM) architecture, we employ a four-layer Transformer architecture with a hidden size of 384. This neural model generates logit values indicating speaker existence. Our diarization model is trained on approximately 3,000 hours of simulated audio mixture data. This data is sourced from the same multi-speaker data simulator used in VAD model training, drawing from the VoxCeleb1\&2~\cite{nagrani2017voxceleb,chung2018voxceleb2} and LibriSpeech~\cite{panayotov2015librispeech} datasets. Additionally, MUSAN~\cite{snyder2015musan} noise is also used for adding additive background noise, primarily focusing on music and broadband noise.

\begin{figure}[t]
\centering
\includegraphics[width=0.4\textwidth]{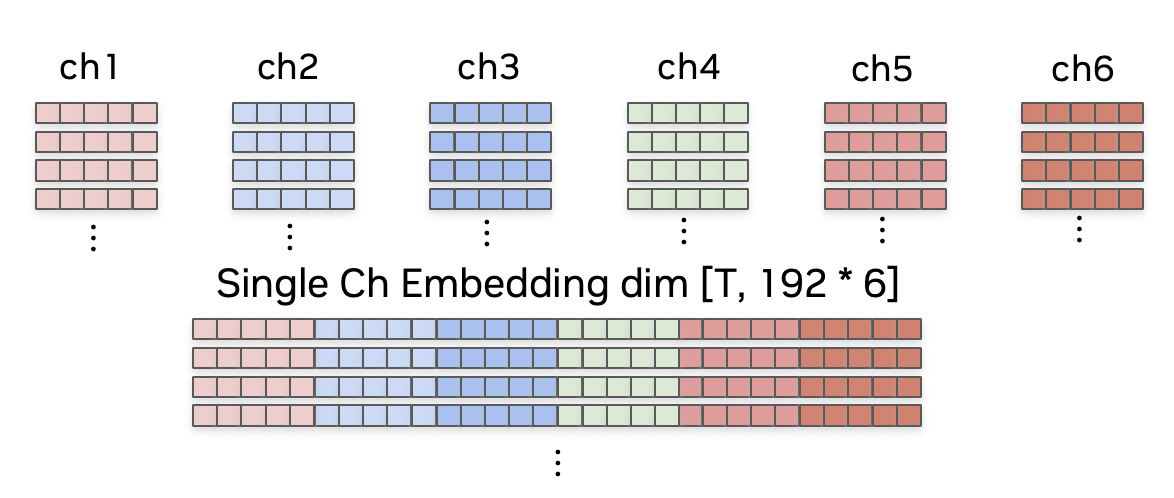}
\vspace{-6px}
\caption{Multi-channel speaker embedding vectors are concatenated to form an multi-channel super-vector that incorporates speaker traits from all channels.}
\label{fig:mc_spk_embs}
\end{figure}

\begin{figure}[t]
\centering
\includegraphics[width=0.45\textwidth]{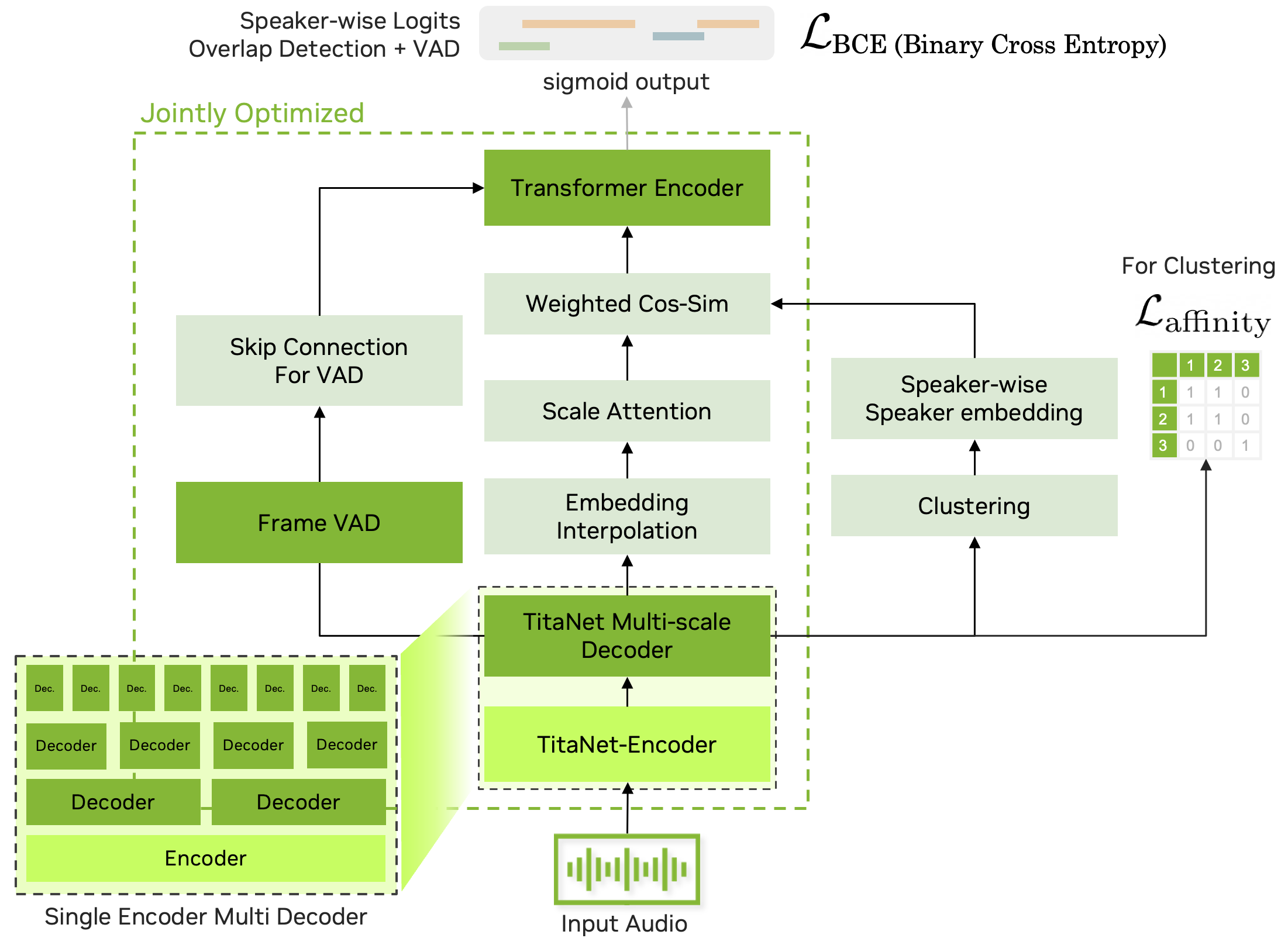}
\vspace{1px}
\caption{Data-flow of speaker diarization system.}
\vspace{3px}
\label{fig:mc_diar_dataflow}
\end{figure}

During inference, we employ NME-SC~\cite{park_2020} for initial clustering. To leverage the multi-channel input, we concatenate the embedding vectors from each channel to create an elongated embedding vector. Specifically, given $M$ clustered channels, the channel displaying the lowest correlation in the speaker embedding vector is excluded, resulting in $(M-1)$ concatenated embedding vectors.
The global speaker clustering result is applied to each channel to produce separate channel-wise diarization outcomes. For multi-scale embedding extraction, we use scale lengths of 3~s, 1.5~s, and 0.5~s with a half-overlap, applicable for both clustering and MSDD inference. For clustering, we adjust the multi-scale weight using the following equation for the $k$-th scale:
\begin{align}
w(k) = r - \frac{r - 1}{K - 1}k,
\label{eq:r_value}
\end{align}
where $k$ is an integer scale index, $K$ denotes the number of scales ($K$=$3$) and $r$ corresponds to the  \texttt{r\_value}. This \texttt{r\_value} is parameterized to define the scale weights using a single floating-point number. We optimize the multi-scale weights on the development set by tuning the $r$ value. If $r>1$, more weight is placed on the longer scale, whereas if $r<1$ the longer scales are given less weight.

We employ a scale interpolation technique on the smallest scale, achieving a finer speaker-decision resolution of 0.05 seconds. As illustrated in Fig.~\ref{fig:frame_vad_mask}, we select the two segments closest to the center of the interpolated segments and then compute the distance to the center of the neighboring segments. The weights for interpolating the scale to obtain the interpolated embedding vector, \textbf{e}, are determined as follows:
\begin{align}
    d_L = d(t_C, t_L) &= |t_C - t_L| \\
    d_R = d(t_C, t_R) &= |t_C - t_R| \\
    \textbf{e} =  \frac{d_L}{\sqrt{d_L^2 + d_R^2}} \textbf{e}_{L}  &+  \frac{d_R}{\sqrt{d_L^2 + d_R^2}} \textbf{e}_{R},
\end{align}
where $t_L$ and $t_R$ represent the timestamps of the centers of the two closest segments, while $\textbf{e}_L$ and  $\textbf{e}_R$ are the two closest speaker embedding vectors on the left and right, respectively. The process of computing interpolated embeddings is executed through a series of matrix-based algebraic multiplications that support batch processing.
The MSDD inference window has a local window length $T_l$ of 15\,s with a hop length of 3\,s. At each inference window, a set of average of speaker embedding vectors is calculated by mixing both the local context (a few tens of seconds, represented by $T_l$) and global context (a few minutes, represented by $T_g$). 
We parameterize the average speaker embedding by using the \texttt{global\_average\_mix\_ratio}, $\alpha$ and the \texttt{global\_average\_window\_length}, $T_g$.~The average speaker embedding vector of each speaker $\textbf{E}_S$, is given by: 
\begin{align}
    \textbf{E}_{S} = \alpha \textbf{E}_{\text{local}, T_{l}} + (1-\alpha) \textbf{E}_{\text{global}, T_{g}},
    \label{eq:mix_ratio}
\end{align}
where $\textbf{E}_{\text{local}}$ and $\textbf{E}_{\text{global}}$ represent the local and global speaker-profile embedding vectors, respectively.~Both vectors have dimension of \texttt{(number of scales,\;embedding dimension,\;number of speakers)}.
The remainder of the training routine is consistent with the system proposed in~\cite{park2022multi}.

\begin{figure}[t]
\centering
\includegraphics[width=0.44\textwidth]{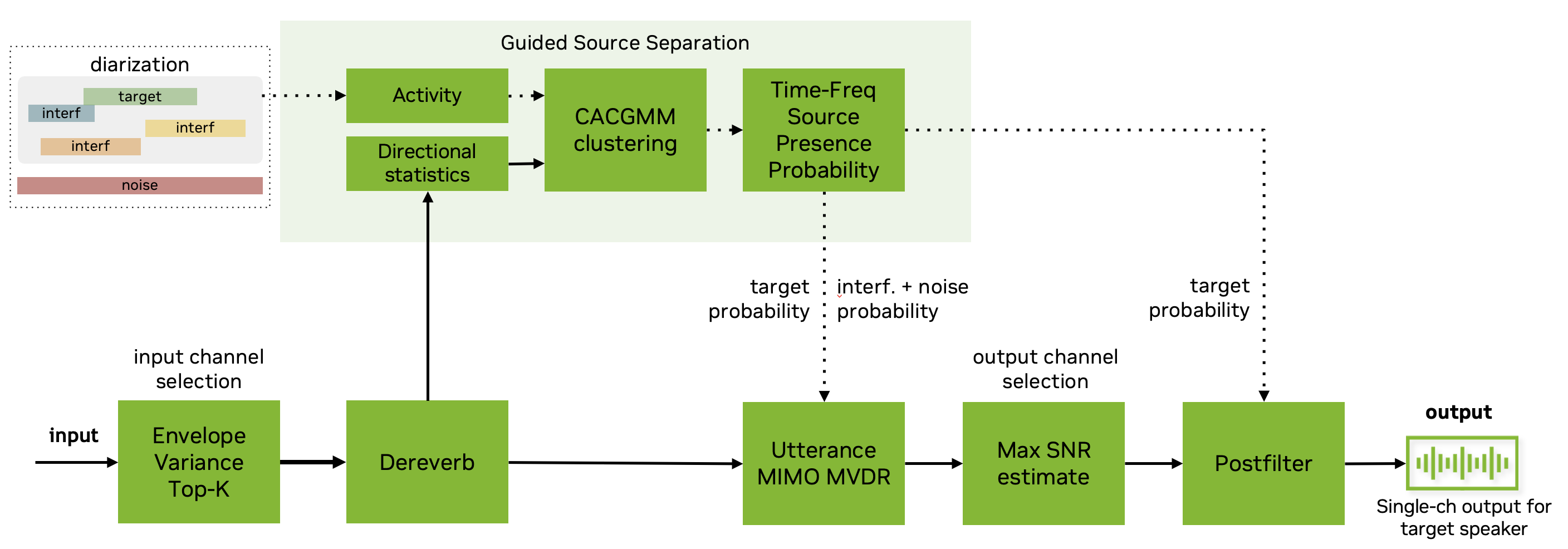}
\vspace{-3px}
\caption{Multi-channel ASR front-end.}
\vspace{-3px}
\label{fig:mc_to_1ch}
\end{figure}

\begin{figure}[t]
\centering
\includegraphics[width=0.44\textwidth]{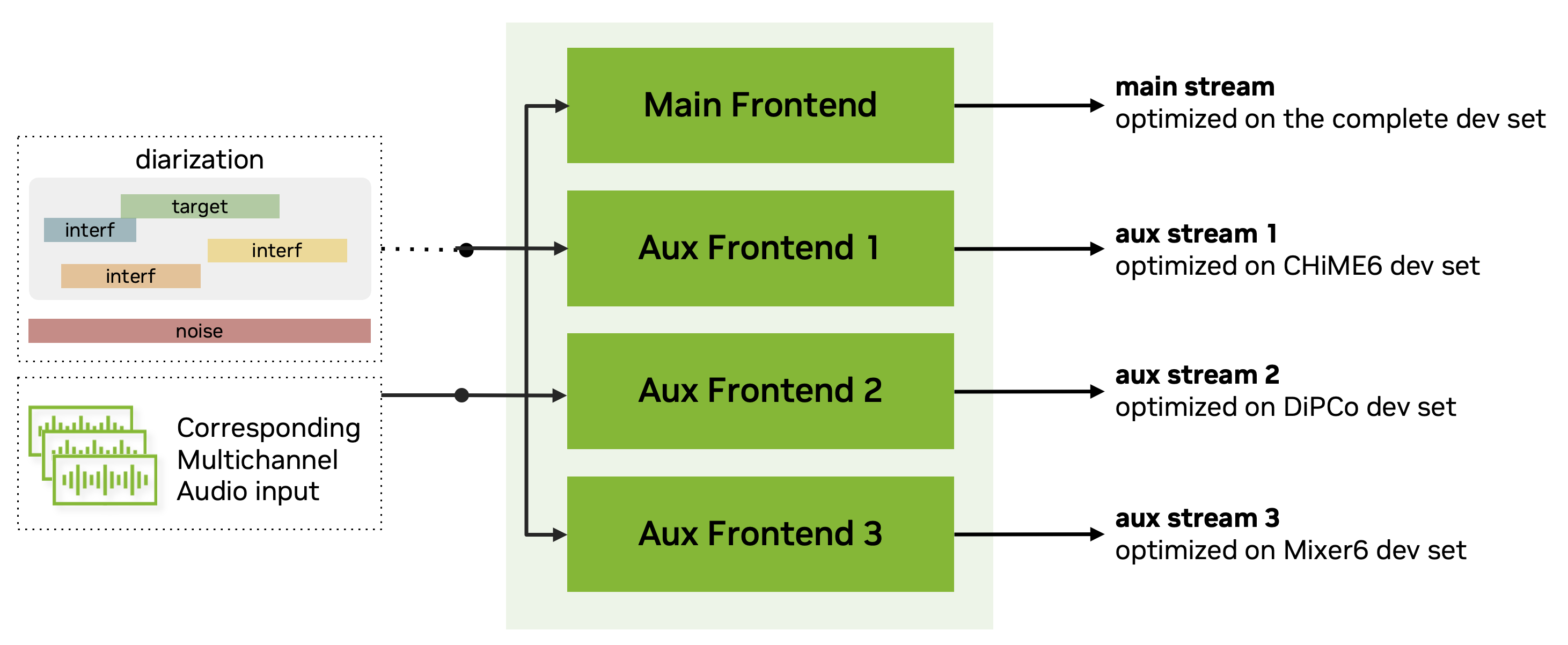}
\vspace{-3px}
\caption{Multi-stream ASR front-end: multiple audio outputs are generated for each input utterance.}
\vspace{-3px}
\label{fig:ensemble}
\end{figure}

\subsubsection{Post-processing of diarization segments}

\begin{figure}[t]
\centering
\includegraphics[width=0.4\textwidth]{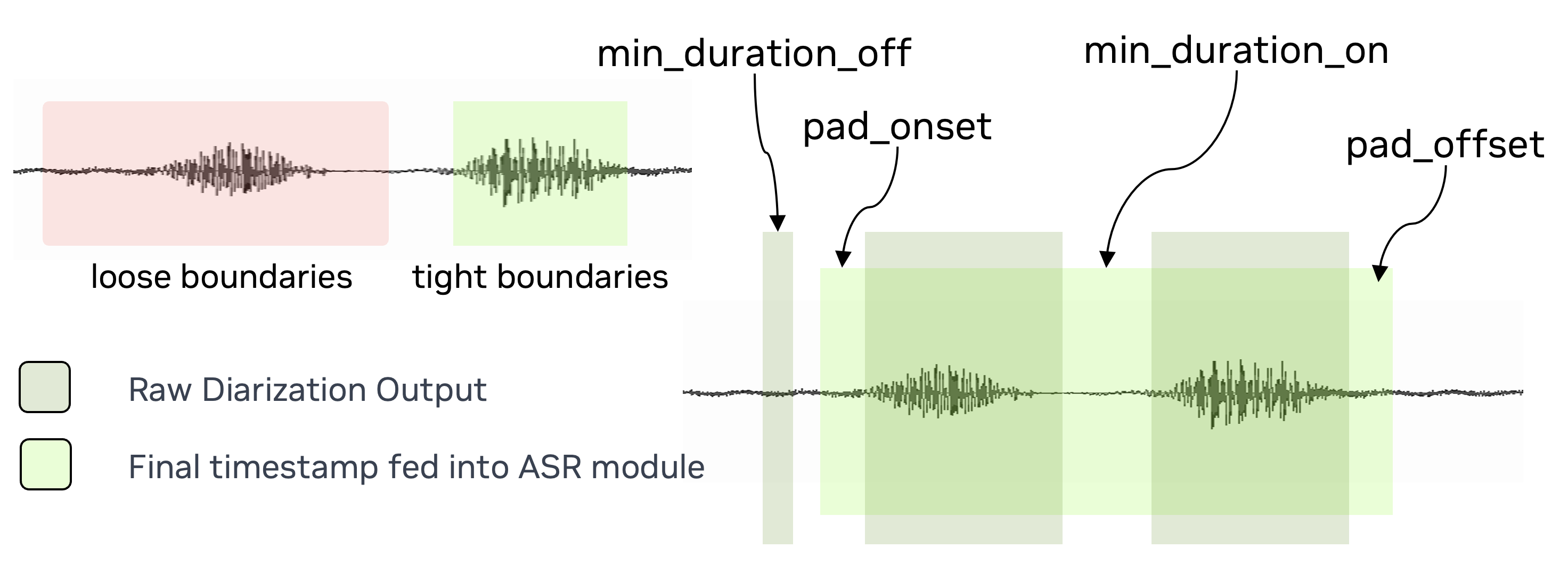}
\vspace{-6px}
\caption{Diarization post-processing using minimum duration, maximum duration, onset padding and offset padding.}
\vspace{3px}
\label{fig:pad_dur_params}
\end{figure}
The final output of the diarization system depicted in \ref{fig:mc_diar_dataflow} is $T \times N_S$ matrix $\textbf{P}_S$ filled with sigmoid values. Here, $N_S$ represents the maximum number of speakers. In our submitted system, we set $N_S$=1. The sigmoid values in $\textbf{P}_S$ are then thresholded using the  \texttt{sigmoid\_threshold}, denoted as $\tau$:
\begin{align}
P_{S}[s, i] > \tau,
\label{eq:sig_thres}
\end{align}
where $s$ and $i$ are indices for the speaker and the finest-scale segments, respectively. The final diarization segments are derived from the timestamps generated by this thresholding process. Finally, we use majority voting on the channel-specific sigmoid values to generate a single set of diarization segments for each speaker.

As previously demonstrated in \cite{medennikov2020stc}, the post-processing of diarization segments can significantly influence the performance of both front-end processing and ASR. Thus, we parameterized the post-processing of these diarization segments, as illustrated in Fig.~\ref{fig:pad_dur_params}. If the segment boundaries are set too tightly, there is a risk of cutting off words; conversely, if they are set too loosely, they may inadvertently include words spoken by the wrong speaker. As depicted, \texttt{pad\_onset} and \texttt{pad\_offset} represent the lengths of buffers appended at the start and end of each segment, respectively. Meanwhile, \texttt{min\_duration\_off} and \texttt{min\_duration\_on} are threshold values used for removing short segments and brief silences, respectively.

\subsection{Multi-channel ASR front-end}
\label{ssec:mc-frontend}

A multi-channel front-end depicted in Fig.~\ref{fig:mc_to_1ch} based on guided source separation (GSS) is used to extract the target speech signal for a single utterance, as in the baseline system~\cite{cornell2023chime}.
The multi-channel front-end consists of envelope variance-based channel selection~\cite{envelope_variance_2014, medennikov2020stc}, MIMO dereverberation~\cite{yoshioka_2012} and mask-based MIMO MVDR beamforming~\cite{souden2009optimal} for extraction of the target speech signal.
Target mask is estimated using GSS with an additional context to prevent permutation~\cite{boeddeker2018front}.
The reference output channel is automatically selected based on maximization of the estimated SNR~\cite{boeddeker2018front} and the target mask with a minimum gain threshold is used to mask the output reference channel.
The front-end is implemented in~\cite{nemo_2019} and optimized to reduce computation time to reduce the time required for parameter optimization of the complete system as described in Section~\ref{sec:hyper_parameter_optimization}.

The main processed audio stream is generated using parameter optimized on the complete \mbox{CHiME-7} development set and used for Systems 2 and 3 in Tables~\ref{tab:system_types} and~\ref{tab:system_comparison_main_dev}.
In the multi-stream configuration depicted in Fig.~\ref{fig:ensemble}, we generate three auxiliary processed audio streams using parameters optimized on each of the three subsets in the development set.
In this configuration, each utterance is processed using all four front-end setups in parallel and the optimal processed audio stream for the current utterance is selected using an ASR-based confidence measure.
For our System-1 in Table~\ref{tab:system_types}, we ensemble the four processed streams with the first preference to main processed audio stream.~An auxiliary stream is used if the corresponding ASR segment confidence is greater than the confidence of the main stream by a threshold, which is tuned to achieve the best performance on the development set.
The ASR confidence is calculated using the method proposed in~\cite{laptev2023fast} with exponentially normalized Tsallis entropy with a temperature of 0.25 and mean aggregation over output tokens. 

\begin{table}[t]
\footnotesize
\caption{System types and descriptions for main track.}
\label{tab:system_types}
\vspace{-8px}
\centering
\begin{tabularx}{\columnwidth}{l X}
\toprule
\textbf{System} & \textbf{Description} \\
\midrule
System-1 & MC Diarization (cascaded optimization) + Multi-stream Front-End Ensemble + ASR + LM \\
System-2 & MC Diarization + Front-End + ASR + LM (end-to-end optimization)\\
System-3 & MC Diarization (cascaded optimization) + Front-End + ASR-LM parameter optimization \\
\bottomrule
\end{tabularx}
\end{table}

\subsection{Automatic speech recognition}
\label{ssec:asr}
Single-channel audio generated using multi-channel front-end (described in Section~\ref{ssec:mc-frontend}) is transcribed using a 0.6B parameter Conformer-based transducer model~\cite{gulati2020conformer}. 
The model was initialized using a publicly available NeMo checkpoint~\cite{conformer_transducer_xl}. It was then fine-tuned on the \mbox{CHiME-7} train set (which includes the \mbox{CHiME-6} and Mixer6 training subsets) after processing the data through the multi-channel ASR front-end, utilizing ground-truth diarization. This fine-tuning phase lasted for 35,000 updates with a batch size set to 128.

Note that we use the ASR model trained only on \mbox{CHiME-7} train set for System-A in Table~\ref{tab:system_comparison_sub_dev} and all three systems in Table~\ref{tab:system_comparison_main_dev}.
Additionally, we also included dev subset data to fine-tune the model used for submission for System-B and System-C in Table~\ref{tab:system_comparison_sub_dev}.
All systems use an external language model as described in Section~\ref{ssec:lm}.

\subsection{N-gram language model}
\label{ssec:lm}
The performance of our end-to-end automatic speech recognition model is improved using beam-search algorithm with a language model (LM).
We apply a word-piece level N-gram language model with byte-pair-encoding (BPE) tokens using SentencePiece~\cite{kudo-richardson-2018-sentencepiece, sentencepiece} and KenLM~\cite{heafield-2011-kenlm, kenlm} toolkits from transcription of \mbox{CHiME-7} train and dev sets.
Token sets of our ASR and LM model were matched.
To combine several N-gram models with equal weights we used OpenGrm library~\cite{roark-etal-2012-opengrm, ngram_library}.
MAES decoding~\cite{kim2020accelerating} was used for transducer.
As expected, the end-to-end model with beam-search decoder and language model performs better than the pure one.
\subsection{Text Normalization}
We implement basic text normalization rules that correct consistent errors or undesired tokens found in the transcript. We remove all double spaces, unknown symbol \textbackslash\texttt{u2047} from ASR model and replace \texttt{aw} with \texttt{oh}. These string mappings are validated using the development set. In addition, we employ \mbox{CHiME-7} scoring for text normalization to produce the final transcription output intended for submission.

\section{Hyper-parameter Optimization}
\label{sec:hyper_parameter_optimization}

\begin{figure}[t]
\centering
\includegraphics[width=0.45\textwidth]{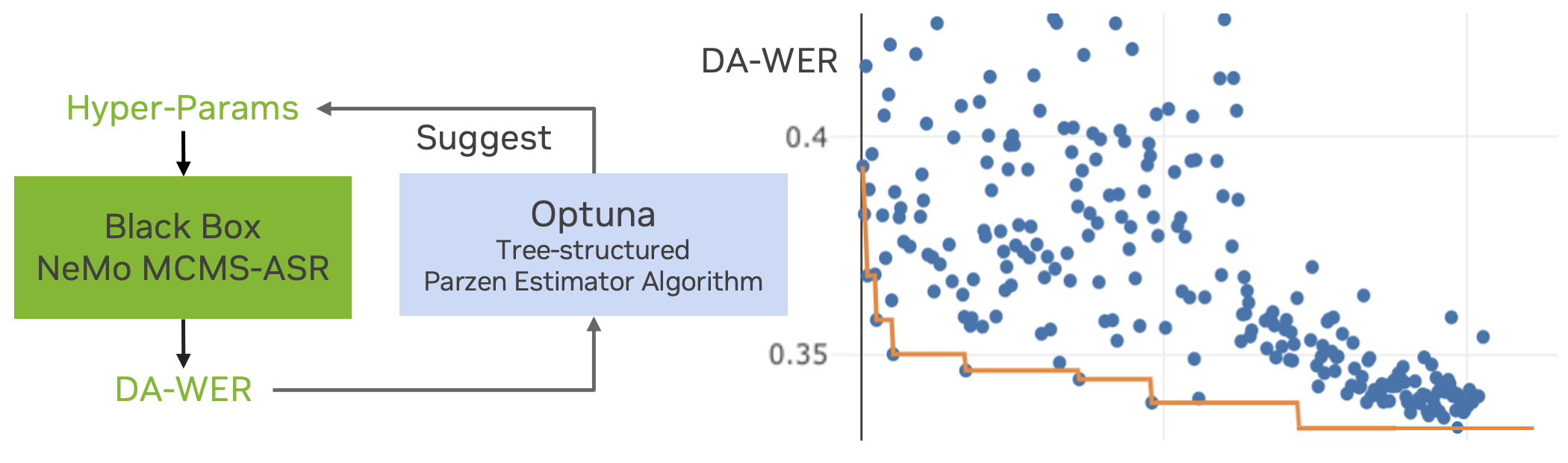}
\vspace{-6px}
\caption{Optuna optimization loop and scatter plot of trials and DA-WER.}
\vspace{3px}
\label{fig:optuna}
\end{figure}

For hyper-parameter optimization, we utilize the \emph{Optuna} framework~\cite{optuna_2019}. This framework facilitates the optimization of parameters for black-box systems based on a target metric that assesses system performance. We employ the default optimization algorithm, the tree-structured Parzen estimator~\cite{ozaki2020multiobjective}, aiming to optimize the macro DA-WER on the development set. As illustrated in Fig.~\ref{fig:optuna}, our DASR pipeline is treated as a black box, given that the optimization does not leverage any prior understanding of the mechanism behind DA-WER computation.
We employ two distinct strategies for system optimization: (1) \emph{cascaded} and (2) \emph{end-to-end} optimization. All optimization tasks run using NVIDIA V100 GPUs, and we run five instances of inference sessions on an 8-GPU machine. The entire process takes roughly 9 hours to produce transcriptions for the complete \mbox{CHiME-7} dev set. Overall, we conduct approximately 2000 trials to identify the best-performing configurations.

\begin{figure}[t]
\centering
\includegraphics[width=0.45\textwidth]{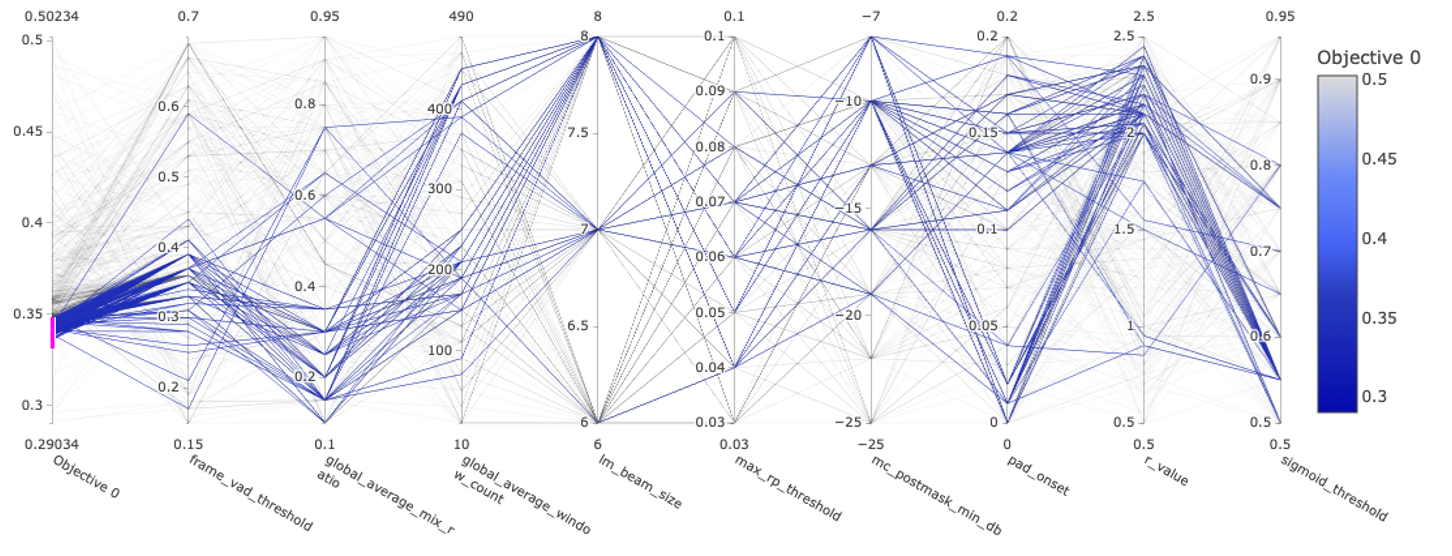}
\vspace{-8px}
\caption{Parallel Coordinate plot of the major hyper-parameters. The systems showed DA-WER in the range of [33.2, 34.5]\% are highlighted.}
\vspace{1px}
\label{fig:optuna_pc}
\end{figure}

\subsection{Cascaded Optimization}
The cascaded optimization comprises two stages. In the first stage, we optimize the diarization front-end, VAD, diarization module, multi-channel ASR front-end, and the ASR module with greedy decoding. In the second stage, we use the front-end outputs as input to an advanced ASR module equipped with a language model and beam-search decoding. At this stage, we focus on optimizing LM scoring and decoding strategies.

\subsection{End-to-End optimization}
The end-to-end optimization tunes the hyper-parameters of all modules simultaneously, encompassing LM rescoring and beam search decoding for ASR. To reduce the search space, we fix certain less important hyper-parameters to the best values identified in the previous cascaded optimization phase and optimize those with greater importance and more varied values across the top-performing trials.

\subsection{Parameter importance}
The hyper-parameters listed below are the top seven that exhibit the most significant importance during the optimization process, sorted in order of their significance:
\begin{itemize}
    \item \texttt{lm\_beam\_size}: beam search width for LM application
    \item \texttt{r\_value}: multi-scale weight scaler $r$ in Eq.~(\ref{eq:r_value})
    \item \texttt{global\_average\_mix\_ratio}: $\alpha$ in Eq.~(\ref{eq:mix_ratio})
    \item \texttt{global\_average\_window\_length}: $T_{g}$ in Eq.~(\ref{eq:mix_ratio})
    \item \texttt{mc\_postmask\_min\_db}: maximum level of post-mask normalization for multi-channel front end 
    \item \texttt{pad\_onset}: pad-onset for diarization output segments
    \item \texttt{sigmoid\_threshold}: sigmoid threshold $\tau$ in Eq.~(\ref{eq:sig_thres})
\end{itemize}
Additionally, Fig.~\ref{fig:optuna_pc} displays a parallel coordinate plot that highlights the relationships among the major hyper-parameters. It is essential to recognize that the concentration of parameter values doesn't directly correlate with parameter importance.

\begin{figure}[t]
\centering
\includegraphics[width=0.44\textwidth]{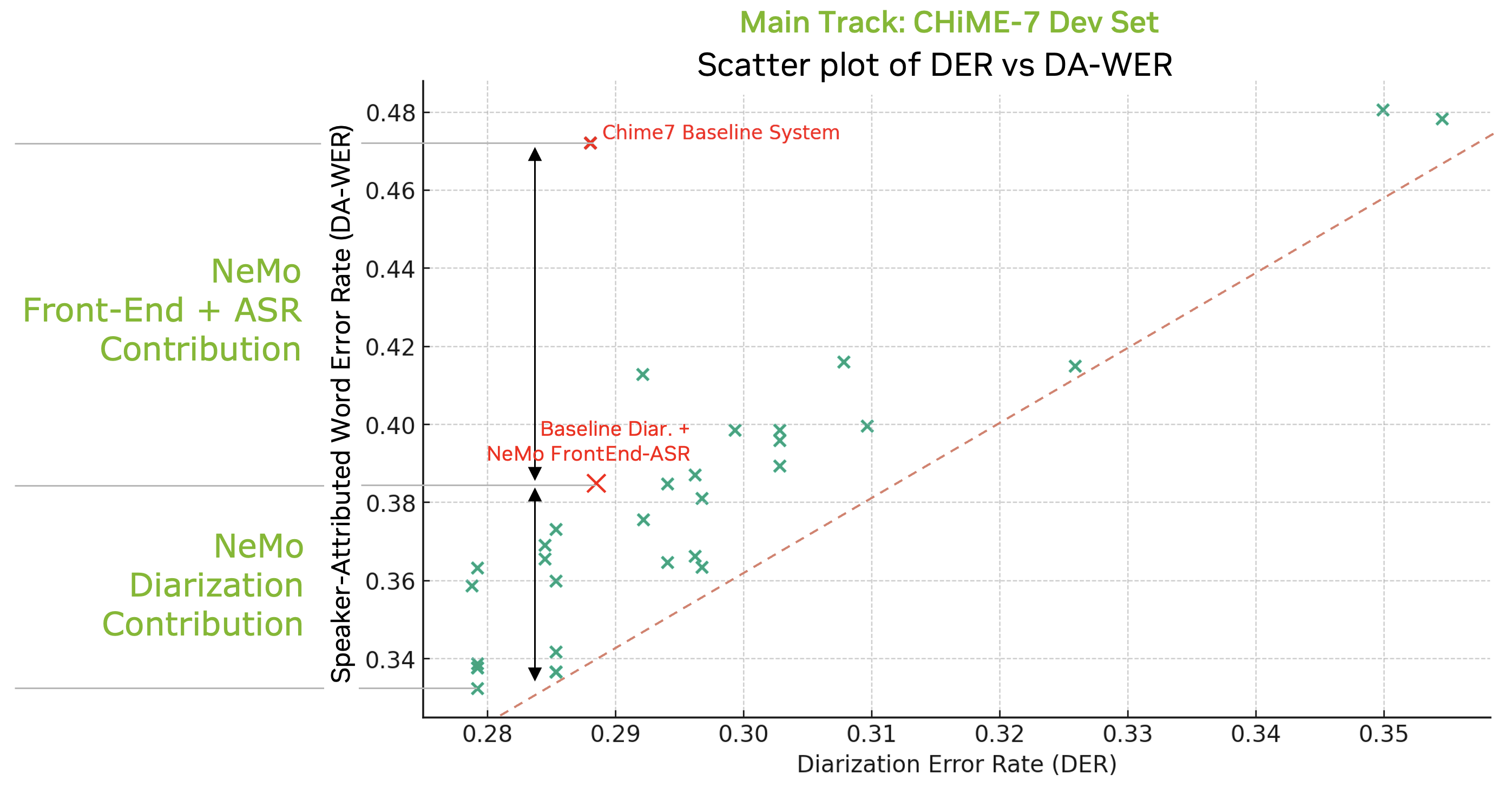}
\vspace{-6px}
\caption{Scatter plots of DER vs DA-WER}
\vspace{-3px}
\label{fig:wer_der_plot}
\end{figure}

\begin{table}[t]
\vspace{4px}
\caption{Sub-track dev set in terms of DA-WER (\%).}
\label{tab:system_comparison_sub_dev}
\vspace{-9px}
\footnotesize
\centering
\begin{tabular}{r|ccc|l}
\toprule
& \textbf{CHiME-6} & \textbf{DiPCo} & \textbf{Mixer6} & \textbf{Macro} \\
\midrule
\textbf{System-A} & 22.8 & 27.6 & 12.8 & 21.0 \\
\textbf{System-B} & 16.9 & 18.7 & 7.9 & 14.4 \\
\textbf{System-C} & 16.7 & 18.8 & 7.9 & 14.5 \\
\bottomrule
\end{tabular}

\vspace{4px}
\caption{Sub-track eval set in terms of DA-WER (\%).}
\label{tab:system_comparison_sub_eval}
\vspace{-9px}
\footnotesize
\centering
\begin{tabular}{r|ccc|l}
\toprule
& \textbf{CHiME-6} & \textbf{DiPCo} & \textbf{Mixer6} & \textbf{Macro} \\
\midrule
\textbf{System-A} & 25.3 & 25.0 & 15.8 & 22.1 \\
\textbf{System-B} & 25.6 & 22.9 & 15.9 & 21.4 \\
\textbf{System-C} & 25.7 & 22.4 & 15.2 & 21.1 \\
\bottomrule
\end{tabular}
\vspace{6px}
\end{table}

For the sub-track, since the oracle diarization segments are provided, we focus on optimizing the ASR (Section~\ref{ssec:asr}) and LM (Section~\ref{ssec:lm}) parameters using Optuna.
All three systems for the sub-task are optimized on audio signals produced using the default front-end parameters of the \mbox{CHiME-7} baseline system.
The results of our three systems for the sub-track are presented in Table~\ref{tab:system_comparison_sub_dev}.
Note that the WER of \mbox{System-B} and \mbox{System-C} are lower than \mbox{System-A} because their ASR model has been trained on the combined \mbox{CHiME-7} train and dev subsets.
The hyperparameters used for ASR models in the sub-task (\mbox{System-A/B/C}) and main task (\mbox{System-1/2/3}) are separately tuned for the two tasks, and do not share the same sets of parameters.

Table~\ref{tab:system_comparison_main_dev} details the systems used in the main track with our diarization system. For System-1 and System-3, we optimize the diarization module, front-end, and ASR simultaneously. Subsequent to this, we select a specific set of parameters for both the diarization and the front-end. For System-3, an ASR model undergoes separate optimization to achieve the lowest DA-WER. For System-2, all modules involved in the DASR task are optimized concurrently through an exhaustive hyperparameter search. It is important to note that the diarization error rate (DER) values in Table~\ref{tab:system_comparison_main_dev} are not necessarily optimal. The systems are primarily tuned to minimize the macro-averaged DA-WER, not the DER.

\section{Experimental Results}
\label{sec:experimental_results}

\subsection{Evaluation Results}
\label{sec:evaluation_results}

Table~\ref{tab:system_comparison_sub_dev} and Table~\ref{tab:system_comparison_sub_eval} display the sub-track results, which are based on the oracle diarization segments. Our best-performing systems exhibit 14.46\% on the dev set and 21.1\% on the eval set. Note that for the dev set result in the sub-track, we rearrange the splits and include the dev set split in the training set for fine-tuning the ASR model. The tendency observed in the dev set persists in the eval set, where dev-set optimized systems (System-B/C) demonstrate improvement over System-A.

Table~\ref{tab:system_comparison_main_dev} and Table~\ref{tab:system_comparison_main_eval} display the main-track results, which are based on the oracle diarization segments. The best-performing systems from our submitted systems score 33.2\% on the dev set and 38.4\% on the eval set. Note that our DER scores are relatively lower than those of other teams or the baseline system because our systems are optimized solely based on the DA-WER value; the DER value is not a factor in selecting the best-performing system. Among our submitted systems, the system with end-to-end hyper-parameter optimization exhibits the lowest DA-WER.

\begin{table}[t]
\caption{Main track dev set in DER, JER and DA-WER(\%).}
\label{tab:system_comparison_main_dev}
\vspace{-8px}
\footnotesize
\centering
\renewcommand{\arraystretch}{0.8} 
\SetTblrInner{rowsep=2pt,colsep=2pt}
\begin{tabular}{rc|ccc|c}
\toprule
& & \textbf{CHiME-6} & \textbf{DiPCo} & \textbf{Mixer6} & \textbf{Macro} \\
\midrule
\multirow{3}{*}{\textbf{System-1}} 
& DER & 38.6 & 29.5 & 15.7 & 27.9 \\
& JER & 39.9 & 32.6 & 19.5 & 30.7 \\
& WER & 41.7 & 40.0 & 18.0 & 33.2 \\ \midrule
\multirow{3}{*}{\textbf{System-2}}
& DER & 41.6 & 28.8 & 17.9 & 29.5 \\
& JER & 43.1 & 31.7 & 21.9 & 32.2 \\
& WER & 42.9 & 39.1 & 18.0 & 33.4 \\ \midrule
\multirow{3}{*}{\textbf{System-3}}
& DER & 38.6 & 29.5 & 15.7 & 27.9 \\
& JER & 39.9 & 32.6 & 19.5 & 30.7 \\
& WER & 42.3 & 40.8 & 18.1 & 33.7 \\
\bottomrule
\end{tabular}
        \begin{tablenotes}
            \item[A]* System-3 uses the same diarization system as System-1.
        \end{tablenotes}
\end{table}

\begin{table}[t]
\vspace{4px}
\caption{Main track eval set in DER, JER and DA-WER(\%).}
\label{tab:system_comparison_main_eval}
\vspace{-8px}
\footnotesize
\centering
\renewcommand{\arraystretch}{0.8} 
\SetTblrInner{rowsep=2pt,colsep=2pt}
\begin{tabular}{rc|ccc|c}
\toprule
& & \textbf{CHiME-6} & \textbf{DiPCo} & \textbf{Mixer6} & \textbf{Macro} \\
\midrule
\multirow{3}{*}{\textbf{System-1}}
& DER & 56.1 & 28.7 & 18.0 & 34.3 \\
& JER & 56.7 & 35.7 & 17.8 & 36.7 \\
& WER & 53.1 & 34.5 & 28.0 & 38.6 \\ \midrule
\multirow{3}{*}{\textbf{System-2}}
& DER & 56.0 & 30.1 & 20.2 & 35.4 \\
& JER & 55.7 & 35.6 & 17.1 & 36.1 \\
& WER & 52.8 & 36.6 & 25.9 & 38.4 \\ \midrule
\multirow{3}{*}{\textbf{System-3}}
& DER & 53.5 & 26.8 & 15.5 & 31.9 \\
& JER & 57.3 & 35.4 & 22.1 & 38.3 \\
& WER & 54.1 & 35.5 & 30.3 & 40.0 \\
\bottomrule
\end{tabular}
\end{table}

\subsection{Ablation study with baseline system}
\label{sec:ablation_study_with_baseline}

As mentioned in the \mbox{CHiME-7} challenge documentation~\cite{cornell2023chime}, the best-performing DA-WER of the baseline system registers at 47.2\%, and the DER at 28.8\%, on the CHiME-7 \textit{Dev} set. We substitute the diarization of our system with that of the baseline diarization system and assess its performance. We refer to this system as \texttt{Baseline Diar.$+$NeMo FrontEnd-ASR}. By evaluating the performance of this system, we discern the contributions of diarization and the other components of the system, such as the Front-end, ASR, LM, and so on, as depicted in Fig.~\ref{fig:wer_der_plot}.
In Fig.~\ref{fig:wer_der_plot}, we present a scatter plot depicting the relationship between WER and DER for all the systems we test during the development process. By substituting the baseline diarization with our diarization system, we enhance the DA-WER from 38.4 to 33.22. It is noteworthy that the DER of our best-performing system for \textit{Dev} stands at 27.86\%, which is close to the 28.8\% of the baseline diarization. However, by utilizing our diarization system, we achieve a significant absolute improvement in DA-WER by 5.2\%. We surmise that a contributing factor to this improvement is that the baseline diarization system exhibits a considerably higher confusion error rate compared to our system's diarization. This discrepancy can potentially lead to doubling the WER count, as a word counted as a deletion in one speaker's transcription might be counted as an addition in another speaker's transcription.

\subsection{Discussions}
The challenge of domain diversity underscores the difficulty of identifying a single optimal configuration due to the unique acoustic and lexical characteristics inherent in each domain. This challenge is further exacerbated by the scarcity of domain-specific training data as our submitted systems are not directly trained on the \mbox{CHiME-6} development set. When considering the multi-channel front-end, it is essential that processing be seamlessly integrated with diarization. This conclusion is derived from the result we get from end-to-end hyper-parameter optimization.~In the realm of ASR techniques, confidence-based ASR ensembling has shown significant promise, although it is not included in the best-performing system. Lastly, pure hyper-parameter optimization alone resulted in over a 16\% relative DA-WER reduction. It is crucial to note that a low DER does not necessarily guarantee a low DA-WER. Conversely, a high DER will never result in a low DA-WER. Moreover, even when DER values are similar, the resulting DA-WER can differ significantly. This variation is evident when comparing false alarms and confusion errors between the baseline diarization and the diarization system of our submitted systems.

\section{Conclusions}
\label{sec:conclusions}
In this paper, we detailed the DASR system of NeMo team for CHiME-7 challange.~Our approach enhances the multi-channel speaker diarization and ASR system with dereverberation and channel clustering, allowing for effective handling of the multi-channel input signal. We then employ late-fusion techniques tailored for multi-channel diarization. Furthermore, to boost the performance of the GSS module, we incorporate MIMO dereverberation and MVDR beamforming techniques. Importantly, we optimize all non-differentiable parameters based on DA-WER, ensuring end-to-end inclusion of all modules. All models and codebases discussed in this paper will be made publicly available.
Future work can encompass various avenues for system enhancement. First, we aim to explore improved methods to adapt the diarization model, specifically MSDD-based models, to domain-specific data. To counteract overfitting on the limited development dataset, we consider employing a mix of regularization techniques, varied training data, and tailored learning rate controls. Second, we intend to further fine-tune the ASR front-end system to cater not only to spatial cues but also to cues related to speaker identity.~We observe that there is potential for refinement in spatial separation, as it occasionally includes speech from the incorrect speaker. Lastly, implementing large language models (LLMs) on top of the decoded ASR output may lead to further reductions in DA-WER by rectifying lexically straightforward errors.


\bibliographystyle{IEEEtran}
\bibliography{mybib}

\end{document}